# Dynamic residential load scheduling based on an adaptive consumption level pricing scheme

Haider Tarish Haider[a,b*], Ong Hang See[a], W. Elmenreich[c]

[a] Department of Electronics and Communication Engineering, Universiti Tenaga Nasional, Jalan IKRAM-UNITEN, 43000 Kajang, Selangor, Malaysia

[b] Department of Computer and Software Engineering, University of Al-Mustansiriyah, 10001 Baghdad, Iraq,

[c] Institute of Networked & Embedded Systems/Lakeside Labs, Alpen-Adria-Universität Klagenfurt, 9020 Klagenfurt, Austria

haiderth@ymail.com; ong@uniten.edu.my; wilfried.elmenreich@aau.at

### Abstract

Demand response (DR) for smart grids, which intends to balance the required power demand with the available supply resources, has been gaining widespread attention. The growing demand for electricity has presented new opportunities for residential load scheduling systems to improve energy consumption by shifting or curtailing the demand required with respect to price change or emergency cases. In this paper, a dynamic residential load scheduling system (DRLS) is proposed for optimal scheduling of household appliances on the basis of an adaptive consumption level (CL) pricing scheme (ACLPS). The proposed load scheduling system encourages customers to manage their energy consumption within the allowable consumption allowance (CA) of the proposed DR pricing scheme to achieve lower energy bills. Simulation results show that employing the proposed DRLS system benefits the customers by reducing their energy bill and the utility companies by decreasing the peak load of the aggregated load demand. For a given case study, the proposed residential load scheduling system based on ACLPS allows customers to reduce their energy bills by up to 53% and to decrease the peak load by up to 35%.

### Nomenclature

*Abbreviations*

| | |
|---|---|
| DR | Demand response |
| DRLS | Dynamic residential load scheduling system |
| CL | Consumption level |
| ACLPS | Adaptive consumption level pricing scheme |
| CA | Consumption allowance |
| HAN | Home area network |
| ToU | Time of use |
| DLC | Direct load control |
| PIB | Price-invariant band |
| CA+ | Positive consumption allowance |
| CA– | Negative consumption allowance |
| BC | British Columbia |
| PR | Price rate |
| R | Rand, currency of South African |

*Variables*

| | |
|---|---|
| $d^t$ | The total load at each time slot |
| t | Time slot |
| $d_p^t$ | One-slot energy consumption for appliance $p$ |
| $D^t$ | The total load of all appliances at time slot $t$ |
| $C_t$ | The pricing function |
| $a_t, b_t, c_t$ | Parameters of quadratic cost function |
| $L_t$ | The customer consumption level in each time slot |
| $S_t$ | The starting operation time of an appliance |
| $E_t$ | The ending operation time of an appliance |
| $OS_t$ | The optimal starting time of appliance operation |
| $OE_t$ | The optimal ending time of appliance operation |

*Parameters*

| | |
|---|---|
| T | Total number of time slots |
| P | Set of household appliances |
| p | One household appliance |
| $d_p$ | The total energy consumption of appliance's p |
| $r_1, r_2, r_3$ | The constant parameters of the ACLPS cost function |
| B | The daily energy cost |
| $B_{Ds}$ | Desired daily customer energy consumption cost |
| $B_{AC}$ | Actual daily customer energy consumption cost |
| Ex | Extended energy consumption |
| $T_D$ | Time duration |

### 1. Introduction

Currently, residential electricity demand accounts for 20% to 40% of the total electrical energy used all over the world [1-3]. Residential loads often contribute significantly to seasonal and daily peak demands [4]. To meet these occasional peak demands, utility companies have been required to increase their generation capacity to match the required demand at alltimes. Generally, about 20% of the power generation capacity is latently available to meet the peak demand that occurs for approximately 5% of the time [5,6]. However, this capacity level is becoming less practical because of the cost of new power plants and the level of greenhouse gas emissions [7,8]. Managing the peak power consumption helps drive significant energy conservation by shifting or curtailing the peak load to achieve smooth customer energy usage [9,10]. Both, utility companies and customers benefit from achieving optimal load management during peak periods [11]. Furthermore, residential homes are becoming smarter because of the integration of the information and communication technologies to connect all household appliances and sensors in a home area network (HAN) for easier monitoring and intelligent control. Meanwhile, smart homes are being faced with varied pricing tariffs where flexible DR schemes are being implemented in many countries all around the world. Therefore, a great opportunity for the residential sector to improve its energy usage load scheduling through smart home techniques under flexible pricing schemes [12–15]. However, residential customers cannot be expected to invest time and obtain knowledge to manage all the smart home devices on their own. Thus, a dynamic load scheduling system is expected to help customers arrange the load scheduling optimally to save energy and cost [16,17].

Recent literature includes several studies that refer to the need to address customer load scheduling in DR systems. In [18], a cooperative game theory model was proposed to optimize the peak load by scheduling the customer appliances. A time-of-use (ToU) DR program was used in the study. The results showed that the customer energy cost was reduced by 18%. In [19], an intelligent home appliance scheduling solution was illustrated on the basis of the ToU program. This solution attempted to optimize the customer constraints and the ToU price change of utility companies to obtain a decision-support system for forecasting the electricity demand and to save energy with an efficient appliance scheduling system. However, the simulation result covered only a few types of customer appliances. Furthermore, customer privacy and load synchronization were not addressed in the study. In [20], a scheduling of actual customer load types was presented using a mixed-integer nonlinear optimization model based on the ToU pricing program. The result achieved approximately 25% cost reduction. A dynamic load scheduling system for a smart home during demand response events was proposed in [21]. In this system, load curtailment and scheduling were adapted every minute to ensure adequate comfort levels during peak periods. Load priorities were fed into an optimization module to determine the least important load at each instant. Another dynamic load scheduling system that incorporated both intelligent smart meter and an aggregator that autonomously scheduled the appliances and storage devices, was proposed in [22]. According to the historical data on the operation of customer appliances, the smart meter learns and predicts the power consumption behavior of the appliances to generate the expected appliance scheduling automatically. The average savings attained by the customers were 20.39%. In [23], a multi-objective genetic algorithm was proposed to optimize the time allocation of domestic load operation while minimizing the costs associated with energy purchase and end-user dissatisfaction. The system showed three extreme solutions to address energy purchase cost, end-use dissatisfaction, and compromise solution. The cost reductions of these three solutions were 24%, 22%, and 23%, respectively. In [24], a mathematical formulation for load scheduling was proposed for optimal cost saving considering the electrical vehicle to home discharging capability. The system investigated different time-varying DR programs (ToU, inclining block rate, and a combination of them) to show the effect of these programs on the results. The results indicated approximately 22% cost reduction. In [25], dynamic load scheduling was proposed on the basis of the theory of optimal portfolio selection. The system optimized the load according to the historical data of customer energy consumption to obtain the customer utility for expected price and energy for the next time slot using the optimal portfolio selection theory. The result achieved about 28% cost reduction. In [26], a dynamic-pricing and peak power limiting-based DR strategy with bi-direction electricity utilization for electrical vehicle and energy storage system was proposed. This system achieved a cost reduction of approximately 65%.

These mentioned works generally refer to load modeling and optimization methods to solve customer load scheduling. On the one hand, most of the current studies on load management aim to schedule the customer load based on a price-based DR scheme, specifically for ToU or real-time pricing programs. In a price-based scheme, the customers are offered time-varying rates that reflect the value and cost of electricity at different time periods [27], which means that the price of energy varies for different times in a day and different seasons in a year [28]. The problem arises from the externality effects of the energy usage of a selfish customer that are imposed on the price rate for other customers. Moreover, customers are offered a single price rate for all consumption levels (CLs) in each period. In addition, customers need to be concerned with price changes with respect to time. On the other hand, the load management studies in the literature optimized the load scheduling based on historical data or expected customer consumption limit, which may not be optimal to reduce the energy bill. The methodology proposed in the present paper intends to use dynamic customer load scheduling based on an adaptive CL pricing scheme to achieve optimal load scheduling. This paper first focuses on modeling the residential load according to actual customer preferences in terms of load scheduling. Second, a mathematical formulation for the objective function and constraints is presented based on actual consumption constraints to manage the customer load scheduling optimally for saving energy and cost. Finally, an adaptive consumption level pricing scheme (ACLPS) is introduced as a DR scheme to overcome the externality effect and time constraint of the price-based DR scheme as discussed in Section 2. In addition, the effect of the price-based program and ACLPS are investigated based on the DRLS results. We consider a scenario where the DRLS functionality is deployed inside the smart meter that is connected to not only the utility side, but also to the HAN to achieve optimal management for the customer's appliances. The overall system performance reveals that employing the dynamic residential load scheduling system (DRLS) benefits not only the customers by reducing their energy cost, but also the utility companies by decreasing the peak load of the aggregated load demand.

## 2. SYSTEM MODEL

In this section, a mathematical formulation for the representation of the demand response scheme, energy consumption model, and pricing model is provided. According to these formulated aspects, we formulate an objective function to optimize the customer load scheduling in Section 3.

### 2.1. DEMAND RESPONSE SCHEME

The current residential DR schemes can be categorized into two types, namely, incentive-based and price-based [29]. In an incentive-based scheme, a utility company offers to manage the loads during emergency or peak periods based on a mutual agreement. A popular DR incentive program is direct load control (DLC). In DLC, the utility company can access and control the appliances of customers and receive an incentive payment or bill credit [30]. This approach, which involves direct access to customer premises for on/off operations, is highly invasive. The lack of customer privacy and system scalability are the major drawbacks of DLC and other incentive-based programs [31].

In a price-based scheme, a customer obtains a price discount for load shifting or reduction during peak periods [32]. Utilities use a distributed process of variable-pricing policies. The customers are encouraged to manage their loads individually by either reducing or shifting their energy consumption from peak hours to less congested hours, thereby favoring load balancing. The price of electricity may differ at pre-set times or may vary dynamically according to day, week, or year [31,33]. The customer reacts to the fluctuations in the electricity prices [34]. Examples of this scheme are ToU, critical-peak price, and real-time price [35,36]. In general, energy price changes with time in every program. In ToU pricing, the utility company changes the ToU price rate according to the available power supply and predicted customer demand [37]. Although price-based schemes do not have customer privacy and system scalability problems, offering a price rate for a specific period to all customers of different CLs is unfair to those who already have a normal or low-level consumption. Following price changes in different periods may also be confusing to customers.

A new DR scheme based on the ACLPS has been used in the proposed load scheduling system. This scheme aims to overcome the drawbacks of the two residential DR schemes. The ACLPS is based on two complementary factors, namely, adaptive pricing for the CL of each individual consumer and consumption allowance (CA). The ACLPS provides customers with an adaptive level of energy consumption pricing over different load operation periods and is based on the monitoring of the energy rate of customers. The proposed scheme significantly differs from the current DR schemes. Unlike the ToU, the proposed scheme fully internalizes the desire of each customer for energy. Thus, the scheme overcomes the externality effects of the usage of a selfish customer by maintaining the price rate for other customers. Moreover, customers are offered multiple price rates according to their CLs instead of a single price rate for all CLs. Customers need not be concerned with price changes with respect to time and can thus use energy freely as long as their consumption levels are within the given allowance. Unlike the DLC, the proposed scheme does not share any critical customer information with utility companies. Thus, customer privacy is not violated. The proposed scheme adopts price rate and CA according to two factors, namely, consumption period and CL. Moreover, the utility company provides multiple price rates that correspond to the level of average consumption as shown in . 1. Each level has a CA associated with it. A price-invariant band (PIB) is double the CA or ±CA around the CL. The CA has two types, namely, positive ($CA^+$) and negative ($CA^-$). A positive CA is defined during the normal consumption period (for the utility), in which the constant price rate is given an allowance around the consumer's level of consumption. A negative CA is defined during the peak consumption period (for the utility). It is the amount of reduction required by a consumer to stay at the same price rate. An incentive is awarded if consumer usage is below the CA of the CL, at which point the consumer is in a lower PIB. The utility company can send a notification message to its customers whenever their consumption changes to a new PIB. Fig. 2 shows the flow chart of the proposed DR scheme. The utility company monitors the consumption data of registered customers provided by smart meters. Based on these data, the price rate and CA changes with respect to the consumption level and period (normal or peak) of customer consumption. The electricity price remains at the same rate if the customer consumption is within the given allowance of the current customer consumption level; otherwise, the utility company charges the customer according to the updated price rate for the new consumption level. To forecast customer consumption levels, ACLPS assign an initial consumption level for the first time that customers participate, based on the standard customer consumption of each country.

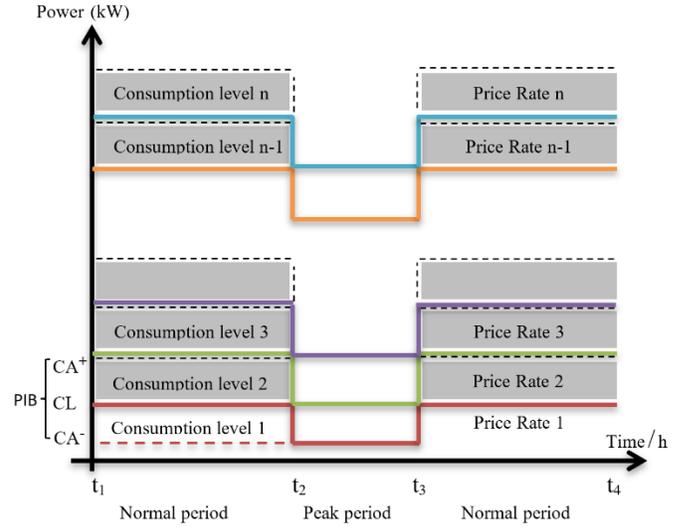

Figure 1 Distribution of energy consumption levels

In accordance with the ACLPS application procedure and based on the assumption that a smart meter is installed, the utility company broadcasts the pricing scheme based on the consumption level. Thereafter, the utility company collects consumption data from the customers. Using these data, the utility company charges the customer and informs its consumers about their typical consumption levels, price rate and the corresponding PIB. Thus, the consumers should manage their energy consumption according to the constraints of the ACLPS to avoid high price rates. Utility companies monitor the consumption levels of customers. The price rate will not change as long as customer consumption is within the PIB region. The increase or decrease in price rate depends on its upward or downward movement into a new PIB region.

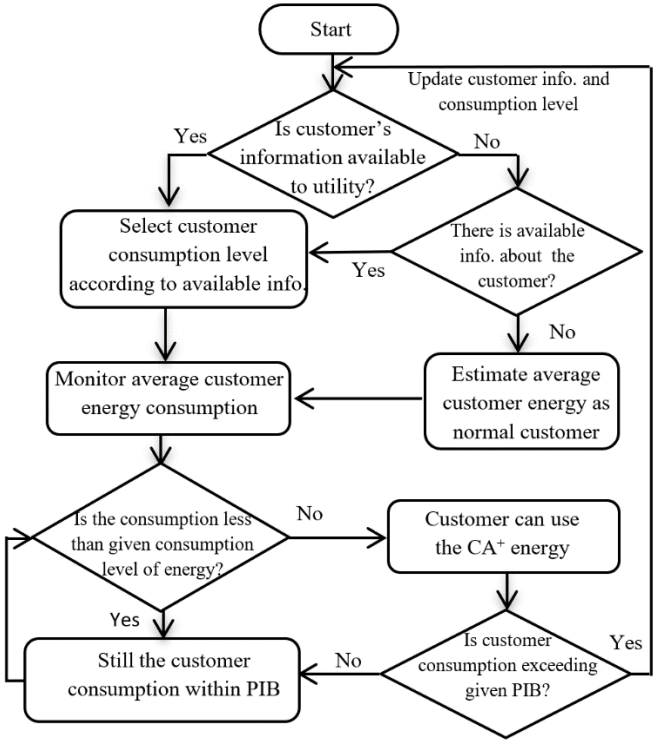

Figure 2 Flow chart of proposed demand response scheme

### 2.2. ENERGY CONSUMPTION MODEL

Each customer is equipped with a smart meter that has DRLS capability to schedule the household energy consumption. The smart meter is connected to the utility company. Additionally, the DRLS system has access to each appliance through the HAN.

Let $d^t$ denote the total load at each time slot $t \in T \triangleq \{1,2,3 \ldots, T\}$. On the basis of a time slot of 5 minutes and a study period of 24 hours, $T = 288$. Let $P$ denote the set of household appliances for a customer, such as air conditioner, washing machine, and refrigerator. For each appliance $p \in P$, the energy consumption can be defined as

$$d_p \triangleq [d_p^1,\ d_p^2, \ldots, d_p^T]. \tag{1}$$

Where $d_p^t$ denotes the corresponding one-slot energy consumption that is scheduled for appliance $p$ at time slot $t$. The total load of all appliances at time slot $t$ is obtained as $D^t = \sum_{p \in P} d_p^t \quad t \in T$. (2)

### 2.3. PRICING MODEL

The pricing function of energy provided by the utility company can be denoted as $C_t(D_t)$ to indicate the cost of customer consumption of $D_t$ units of energy in each time slot $t$. The following assumptions are obtained for the cost function [38].
Assumption 1: The cost functions increase with respect to the total energy consumption such that

$$C_t(\breve{D}_t) < C_t(\tilde{D}_t) \quad \forall\ \breve{D}_t < \tilde{D}_t \tag{3}$$

For (3), energy cost increases if the total load increases.
Assumption 2: The cost functions are strictly convex.
Assumption 3: There exists a convex, non-decreasing function $g_t(s)$ for the domain $s \in [0, +\infty)$ for each $t \in T$, $g_t(0) \geq 0$, and at $s \to \infty$, $g_t(s) \to \infty$ such that

$$g_t(s) = \int_0^s g_t(z)\, dz. \tag{4}$$

One interesting example is the quadratic function that satisfies the above assumptions as a cost function [39] as shown in Fig.3 (a).

$$C_t(D_t) = a_t\, D_t^2 + b_t\, D_t + c_t \tag{5}$$

where $a_t > 0$, $b_t \geq 0$, and $c_t \geq 0$ are fixed parameters at each $t \in T$.
Such pricing tariffs could be used by the utility to impose a proper load control with a price-based scheme. For example, British Columbia (BC) Hydro in Canada adopts a convex price model in form of a two-step piecewise linear function to encourage energy conservation [18] as shown in Fig.3 (b).
According to the ACLPS, the proposed pricing function shall take the effect of customer CL to provide customers a multiple price rate according to the level of consumption unlike time-varying of the price-based DR scheme. Based on the ACLPS, the pricing function can be written as

$$C_{l,t}(D_t) = A_{l,t}\, D_t^2 + B_{l,t} D_t + C_{l,t}.$$
where $A_{l,t} = r_1.L_t,\ B_{l,t} = r_2.L_t\ and\ C_{l,t} = r_3.L_t$ (6)

Where $A_{l,t}, B_{l,t}$ and $C_{l,t}$ are same parameters as in (5) but they change with respect to the customer consumption level $L_t$ in each time slot instead of a time-varying price-based DR scheme. Moreover, $r_1 > 0, r_2 \geq 0$, and $r_3 \geq 0$ are the constant parameters of $A_{l,t}$, $B_{l,t}$ and $C_{l,t}$ respectively selected based on energy cost and utility profit policy; and $L_t$ is the customer CL in each time slot. Therefore, in each time slot $t$, the energy cost changes according to the level of customer consumption $L_t$ and the amount of customer energy consumption $D_t$ in (kWh).

### 3. PROBLEM FORMULATION

An efficient customer load scheduling system can be formulated in terms of minimizing the energy costs for electricity usage, which can be expressed as the following optimization problem such that:

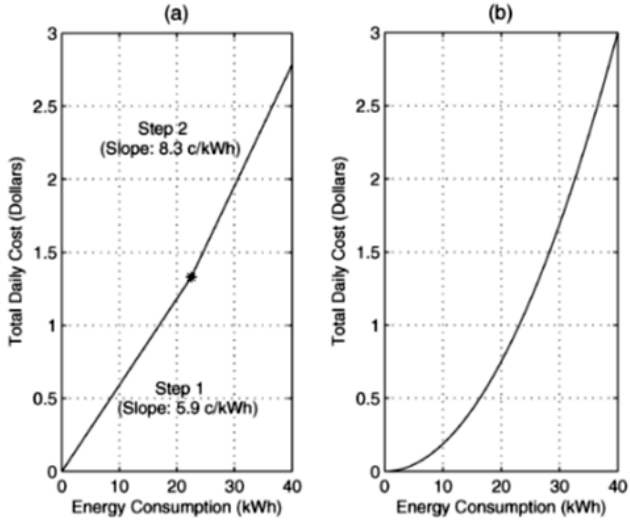

Figure 3 Two sample convex and increasing pricing functions [18]: (a) Two-step conservation rate model used by BC Hydro; (b) A quadratic pricing function.

$$\underset{d \in D}{minimize} \ \sum_{t=1}^{T} C_t \left( \sum_{p \in P} d_p^t \right). \quad (7)$$

To minimize the energy cost for each customer, let $B$ denote the daily energy cost to be charged to the customer by the utility company at the end of each day.

$$B \geq \sum_{t=1}^{T} C_t(D^t). \quad (8)$$

According to the proposed DR scheme, each customer has his/her own PIB, CL, and associated price rate (PR). According to these actual consumption constraints, customers need to optimize their load scheduling so that no further energy is needed over the given CA.
Another assumption is that

$$\frac{B_{Ds}}{B_{Ac}} = \frac{\sum_{t=1}^{T}(D_{Ds}^t)}{\sum_{t=1}^{T}(D_{Ds}^t) + \sum_{t=1}^{T}(D_{Ex}^t)} \quad (9)$$

where, ($Ds$) is the total desired daily customer energy consumption cost and ($Ac$) is the total actual daily customer energy consumption cost which equals the desired energy consumption plus the extended ($Ex$).
Equation (9) can be re-arranged such that

$$B_{Ac} = B_{Ds} \left( 1 + \left( \frac{\sum_{t=1}^{T}(D_{Ex}^t)}{\sum_{t=1}^{T}(D_{Ds}^t)} \right) \right). \quad (10)$$

To optimize equation (10), customers have to make the extended consumption zero so that .

$$B_{Ac} \leq B_{Ds}. \quad (11)$$

$$minimize \left( 1 + \left( \frac{\sum_{t=1}^{T}(D_{Ex}^t)}{\sum_{t=1}^{T}(D_{Ds}^t)} \right) \right). \quad (12)$$

The DRLS system can solve problem (12) for optimal load scheduling as long as utility company provides the customers consumption constraints (PIB, CL and PR) through the smart meter as well as customer provides the initial load scheduling based on the usual or preferred usage. Based on these customer constrains, the DRLS dynamically and automatically schedules the customer load for optimal cost and energy saving. In this work, the appliances are flexible within the customers' specified time ranges. For each appliance, the user indicates $S_t$ and $E_t$ as the beginning and end of the time duration ($T_D$), respectively, within which the appliance use is to be scheduled. $T_D$ is the allowable time interval or time duration required to finish the normal operation of the appliance. The DRLS should help the customer to minimize the extended consumption by solving the problem 12 to reduce customer payment. A flow chart describing the proposed DRLS load scheduling system to address this problem is shown in Fig.4.

## 4. CASE STUDY

The authors of this research encountered difficulties in obtaining actual customer load data in the University Tenaga Nasional Malaysia. Therefore, we adapted the case study of customer load from [20], which studied ten appliances of a typical household in South Africa. The rated power of the appliances, as well as the $S_t$ and $E_t$ information, within which the appliances are to be scheduled, were specified by the customers based on the normal or preferred usage and matched with the obtained usage data. These data on the appliance usage in a household were collected for all weekdays in a month; information on the allowable time duration ($T_D$) that is required to complete the normal operation of the appliance was also obtained and the highest value is recorded in Table 1. Most activities in a typical working class household occurred in the morning and after work. Furthermore, the energy cost function was assumed to be quadratic with the adaptive level consumption based on the ACLPS of (6).

Table 1 shows the customer load parameters used. Appliance 1 (stove) was scheduled to operate twice a day for at least 30 and 50 minutes in the morning and evening, respectively. The DRLS model had a 5-minute sampling time and optimization was beyond a 24-hour period, which encourages shorter waiting periods for behavior changes. By contrast, a 10-minute sampling time was used in [20]. Therefore, a stove should be switched on between the time interval of t=60 (05:00) to t=84 (07:00) and t=192 (16:00) to t=240 (20:00). Appliance 2 (microwave) was scheduled to operate once a day for at least 50 minutes, any time from t=216 (18:00) to t=228 (19:00). The baseline schedule of the appliance was specified by the customer based on the normal or preferred usage.

The parameters of cost function (6) ($r_1$, $r_2$, and $r_3$) were selected to provide the same total utility revenue of approximately R25.37 before scheduling, which was similar to that of [20]. R denotes the South African currency, ZAR or rand. Furthermore, the customer consumption levels $L_t$ starting from 0.02 to 11 kW of step 0.07 were selected to include the highest peak consumption of customer load. In addition, the normal consumption period per day was assumed to be 16 hours, from 01:00 to 7:00, 11:00 to 18:00, and 22:00 to 01:00. Moreover, the peak period was assumed to be 8 hours, from 7:00 to 11:00 and 18:00 to 22:00.

Table (1) The baseline and optimal appliances scheduling

| Load type | S/N o. | Appliances Name | Power rate (W) | Duration (slot/day) | St (slot) | Et (slot) | OSt (slot) | OEt (slot) |
|---|---|---|---|---|---|---|---|---|
| Common no-shiftable | 1 | TV | 100 | 48 | 217 | 264 | - | - |
| | 2 | ceiling Fan | 85 | 288 | 1 | 288 | - | - |
| | 3 | Table Fan | 35 | 72 | 193 | 264 | - | - |
| | 4 | Refrigerator | 72 | 288 | 1 | 288 | - | - |
| | 5 | Telephone, charger | 6 | 288 | 1 | 288 | - | - |
| | 6 | Lighting | 28 | 60 | 217 | 276 | - | - |
| | 7 | Refrigerator (one door normal) | 42 | 288 | 1 | 288 | - | - |
| Selective no-shiftable | 8 | Sound equipment | 5 | 36 | 205 | 240 | - | - |
| | 9 | Printer | 80 | 12 | 241 | 252 | - | - |
| | 10 | Mixer | 120 | 2 | 221 | 222 | - | - |
| | 11 | Bug Killer light | 40 | 12 | 193 | 204 | - | - |
| | 12 | Rice Cooker | 300 | 9 | 217 | 225 | - | - |
| Common shiftable | 13 | Electric Kettle | 1800 | 2 | 199 | 200 | 49 | 50 |
| | 14 | Microwave/Oven | 800 | 2 | 217 | 218 | 205 | 206 |
| | 15 | Dishwasher | 1200 | 31 | 215 | 245 | 97 | 127 |
| | 16 | Water Heater | 2600 | 11 | 210 | 220 | 37 | 47 |
| | 17 | thermos flask | 700 | 1 | 72 | 72 | 49 | 49 |
| | 18 | Air Conditioner 1.5 Horse Power | 995 | 36 | 229 | 264 | 1 | 36 |
| | 19 | Washing Machine (Automatic) Top load | 450 | 10 | 240 | 249 | 97 | 106 |
| Selective shiftable | 20 | Hair Dryer | 1200 | 7 | 66 | 72 | 49 | 55 |
| | 21 | Sandwich Toaster | 700 | 4 | 240 | 243 | 49 | 52 |
| | 22 | Vacuum Cleaner | 600 | 7 | 230 | 236 | 61 | 67 |
| | 23 | Dry Iron | 1000 | 10 | 181 | 190 | 181 | 190 |
| | 24 | Toaster | 700 | 3 | 73 | 75 | 49 | 51 |
| | 25 | Coffee Maker | 700 | 12 | 229 | 240 | 205 | 216 |
| | 26 | Trash compactor | 450 | 3 | 262 | 264 | 181 | 183 |

Finally, we considered that the utility company provides the DRLS system with customer consumption constraints (PIB, CL and PR) of the ACLPS scheme with the use of a smart meter.

### 4.1. SIMULATION AND RESULTS

The proposed formulated model for load scheduling was solved using an iterative optimization program in MATLAB. The simulation results on the optimal and baseline customer load scheduled are shown in Table 2. These results show that customer loads are redistributed from baseline scheduling to different time slots to obtain a significant cost reduction in the energy bill. Based on Table 2, the baseline schedule of a stove has 6 slots (30 minutes) in the morning and 10 slots (50 minutes) in the evening at $S_t$= 73 (06:05) to $E_t$=79 (06:35) and at $S_t$= 212 (17:40) to $E_t$=222 (18:30), respectively. The suggested optimal time scheduling for a stove is at $OS_t$= 77 (06:25) to $OE_t$=83 (06:55) in the morning and at OSt= 203 (16:55) to OEt=212 (17:40) in the evening. The second appliance (microwave), works once a day for at least 2 slots/day (10 minutes) and the baseline and optimal scheduling time is the same starting from slot 216 (18:00) to slot 218 (18:10). Both baseline and optimal solution are the same for the microwave appliance because the baseline operation period is totally within the peak period. The results of the remaining appliances are shown in Table 2; other appliances have remained in the normal period while a few overlap in both peak and normal periods. This result is expected for customers with a wider range of possible start and end operational times of appliances.

The simulation results of total customer energy consumption with and without DRLS based on the ACLPS are shown in Fig. 5. The figure shows that peak customer load is managed so that most loads are shifted to the normal period to maintain low energy bill. According to this optimal load scheduling, the total daily customer energy cost is reduced from R25.37 to R11.76 (i.e., 53%). Furthermore, the peak load is decreased from 10.5 kW at the peak time of 18:20 to 6.83 kW at normal time of 16:00 (i.e., 35%) as indicated in Fig. 5. Table 3 shows the comparison of method and results of DRLS and the model presented in [20]. The proposed system used iterative optimization for dynamic and automatic customer load schedules. Iterative optimization gives all possible loads scheduling solutions within given customer allowable operation time period. Based on these solutions DRLS can select the optimal one that gives significant cost and energy saving. Therefore, DRLS can provide significant cost reduction of 53% compared to 25% in [20]. Furthermore, the peak energy decreases to 35% in DRLS compared to 20% in [20] for same customer data, where a mixed integer non-linear programming using advanced interactive multidimensional modeling system (AIMMS) software has been used to optimize the customer load scheduling. Based on these results, AIMMS in [20] provides a local optimization solution for the given case study. Customer consumes the same amount of energy in the two cases; the difference is the more efficient scheduling of his/her consumption in the case where DRLS is used. The cost reduction changes according to the type of selected appliances and the flexibility of the operation time are provided by the customer for the predefined appliances.

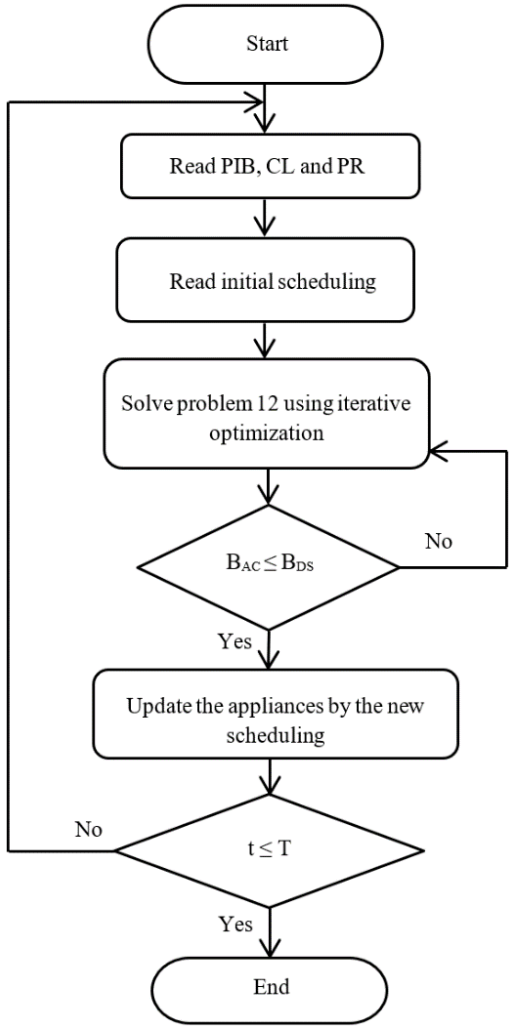

Figure 4 Flow chart illustrated the proposed load scheduling system

### 4.2 VALIDATION OF THE PROPOSED SCHEDULING MODEL

In this subsection, the effect of price-based and ACLPS schemes on the proposed load scheduling are investigated. To highlight the difference between the proposed model and previously published models, a comparison between the proposed scheduling model and the model used in [20] is conducted. In [20], a ToU tariff of 144.52 c/kWh was used for the peak periods (07:00 to 11:00 and 18:00 to 21:00) and a tariff of 45.54 c/kWh was used for normal periods (01:00 to 7:00, 11:00 to 18:00, and 22:00 to 01:00).

Table 4 shows the result of comparing Setlhaolo's scheduling model in [20] and the proposed scheduling model DRLS. For DRLS, two types of tariffs are used, namely, the ToU used in [20] and ACLPS, to determine the effect of the DR scheme on the scheduling model. The total customer consumption, peak level of consumption, and total cost are shown in the results for both cases of before and after load scheduling. Before scheduling, all the cited parameters for customer energy consumption and cost are the same for both models because scheduling is based on baseline customer data. Furthermore, the parameters are included in the results in Table 4 for ease of comparison of results after the scheduling of each model.

Table 4 shows that Setlhaolo's model shifted several appliances from peak to normal to reduce the cost from R25.37 to R18.80 (i.e., 25% cost reduction). Furthermore, peak consumption decreased from 10.5 kW at peak period to 8.4 kW (i.e., 20%) at normal period. For the proposed scheduling model, customer consumption was redistributed to shift the most allowable load from peak to normal period within the given incentive CA. Peak load decreased from 10.5 kW at peak time to 6.8 kW at normal period (i.e., 35%). The total cost was reduced from R25.37 to R17.38 (i.e., 31%) based on the ToU tariff of [20]. Based on the ACLPS DR scheme, the cost was reduced to R11.76 (i.e., 53%). According to these results, DRLS is capable of higher reduction in terms of cost and peak energy consumption compared with the model used in [20]. Furthermore, DRLS-based ACLPS DR scheme provided higher cost reduction than ToU tariff, when the same total utility revenue for both tariffs was assumed before scheduling. In ACLPS, customers receive incentives as long as their consumptions are within the assigned PIB. PIB allows customers to shift the schedule of more appliances to the normal period, whereas in ToU, all customers of different CLs have the same price rate for each time slot. Using DRLS based

Table (2) The customers energy and cost data

| DR scheme | Without DLSS | | | | With DLSS | | | | |
|---|---|---|---|---|---|---|---|---|---|
| | Total energy (kWh) | Total energy in peak period (kWh) | Total energy in normal period (kWh) | Total cost ($) | Total energy (kWh) | Total energy in peak period (kWh) | Total energy in normal period (kWh) | Total cost ($) | Cost reduction (%) |
| ToU | 18.5357 | 11.4636 | 7.0721 | 4.0125 | 18.5357 | 2.9653 | 15.5705 | 2.4202 | 40 |
| ACLPS | 18.5357 | 11.4636 | 7.0721 | 4.0125 | 18.5357 | 2.9653 | 15.5705 | 1.4162 | 65 |

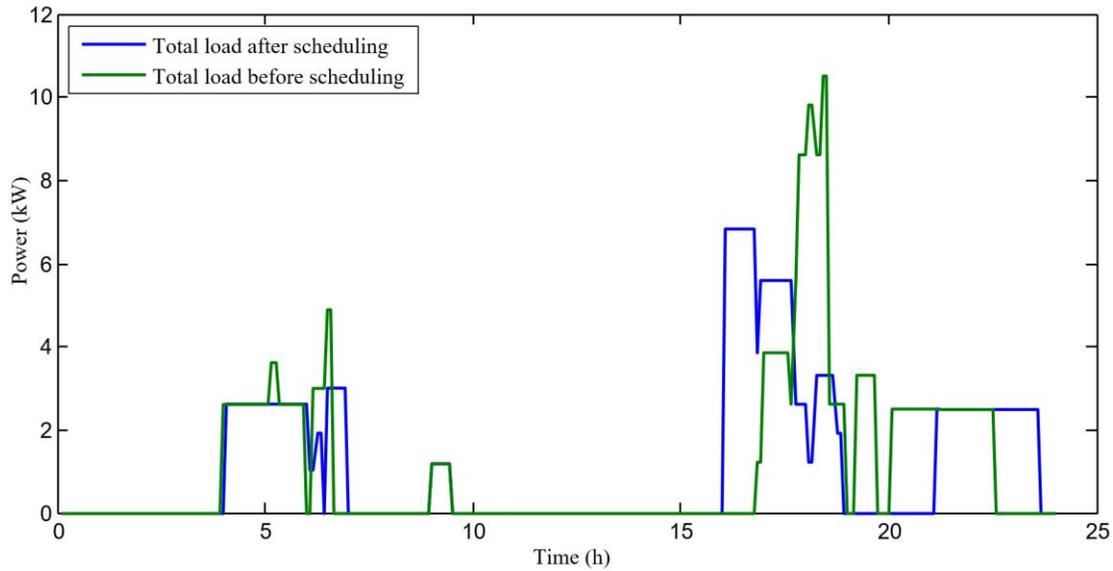

Figure 5 The total customer consumption with and without using DRLS

Table (3) The comparison between Setlhaolo's and proposed models

|  | DR Scheme | Optimization Method | Sampling Time (min.) | Peak consumption (kW) | Total cost (R) | Cost reduction (%) | Peak reduction (%) |
|---|---|---|---|---|---|---|---|
| Setlhaolo's Model | TOU | Mixed integer programming (AIMMS software) | 10 | 8,4 | 18,8 | 25 | 20 |
| DRLS Model | ACLPS | Linear programing (a MATLAB iterative program) | 5 | 6,8 | 11,76 | 53 | 35 |

Table (4) Customers data for energy and cost

| | Before scheduling | | | After scheduling | | | | |
|---|---|---|---|---|---|---|---|---|
| DR Scheduling model | Total energy (kWh) | Peak consumption (kW) | Total cost (R) | Total energy (kWh) | Peak consumption (kW) | Total cost (R) | Cost reduction (%) | Peak reduction (%) |
| Setlhaolo's scheduling-based ToU tariff | 27,18 | 10,5 | 25,37 | 27,18 | 8,4 | 18,8 | 25 | 20 |
| DRLS based ToU tariff | 27,18 | 10,5 | 25,37 | 27,18 | 6,8 | 17,38 | 31 | 35 |
| DRLS based ACLPS tariff | 27,18 | 10,5 | 25,37 | 27,18 | 6,8 | 11,76 | 53 | 35 |

on ACLPS instead of ToU can provide customers with higher cost reductions.

### 4.3. DISCUSSION

The results indicate that the DRLS can maintain significant cost reduction for the customer and peak reduction for the utility. The proposed customer load scheduling actively encourages customers to be aware of their energy consumption so that they can limit it within the given allowance. In addition, the ACLPS adapts the price rate according to the average customer consumption. Customers are awarded with incentive consumption allowances during normal periods so that DRLS can help customers shift the operation of more appliances optimally from the peak period to the normal period to reduce energy bills. In this method with CA, customers need not be concerned about the impact of the energy consumption of selfish customers, and they have flexible DRs provided by the CA. Given an adaptive DR scheme and dynamic load scheduling system, customers and utility companies have a powerful and flexible tool to manage the available generation capacity. Both the utility companies and customers are expected to benefit from the proposed DRLS as a means to balance the available generation capacity and minimize energy cost effectively.

### 5. CONCLUSION

In this paper, a dynamic load scheduling system was proposed on the basis of an adaptive CL pricing scheme to minimize the cost of energy and balance the customer consumption in different periods. Unlike most of the previous studies that focused on scheduling the customer load on the basis of the time-varying price-based DR scheme and historical or expected consumption limits, the present study focuses on the load management based on actual consumption constraints (PIB, CL and PR) provided by utility. Furthermore, the mutual independence of the energy consumption costs among consumers based on the ACLPS. The proposed DRLS optimally schedules the customer appliances on the basis of a mathematical objective function according to typical customer consumption constrains in DR scheme. In addition, ACLPS prevents the selfish customer from imposing the price rate on the other customers. The consumers are charged based on their actual consumption. In this way, customers are encouraged to be aware of their energy CLs to avoid high energy bills, whereas in price-based schemes, all consumers are penalized because of a small number of selfish consumers with high energy consumption. The proposed scheme also provides a consumption band (CA+) during normal periods at a fixed price rate. This provision serves as an incentive for customers to shift their energy consumption from the peak periods to within the CA+ to reduce their energy bills. Therefore, the proposed DRLS helps customers achieve a consumer-targeted energy cost. As consumers pre-plan their consumption levels, the DRLS helps maintain the stability of the power grid through the dynamic load scheduling for shiftable loads. The simulation results indicate that the DRLS based on ACLPS helps customers reduce the cost by as much as 53% compared with the 31% for the ToU-based system, and the peak load is reduced by 35%. The proposed load scheduling is a comprehensive DR solution that benefits both the utility companies and their customers because it enables consumers to manage their energy bills and utility firms to control aggregate consumption levels.

## REFERENCES

[1] Wu Z, Zhou S, Li J, Zhang XP. Real-time scheduling of residential appliances via conditional risk-at-value. IEEE Transactions on Smart Grid 2014;5:1282–91.

[2] Alberini A, Filippini M. Response of residential electricity demand to price: The effect of measurement error. Energy Econ 2011;33:889–95.

[3] Torriti J. A review of time use models of residential electricity demand. Renew Sustain Energy Rev 2014;37:265–72.

[4] He Y, Wang B, Wang J, Xiong W, Xia T. Residential demand response behavior analysis based on Monte Carlo simulation: The case of Yinchuan in China. Energy 2012;47:230–6.

[5] Bassamzadeh N, Ghanem R, Lu S, Kazemitabar SJ. Robust scheduling of smart appliances with uncertain electricity prices in a heterogeneous population. Energy Build 2014;84:537–47.

[6] Strbac G. Demand side management: Benefits and challenges. Energy Policy 2008;36:4419–26.

[7] Zakariazadeh A, Jadid S, Siano P. Economic-environmental energy and reserve scheduling of smart distribution systems: A multiobjective mathematical programming approach. Energy Convers Manag 2014;78:151–64.

[8] Beaudin M, Zareipour H. Home energy management systems: A review of modelling and complexity. Renew Sustain Energy Rev 2015;45:318–35.

[9] Li XH, Hong SH. User-expected price-based demand response algorithm for a home-to-grid system. Energy 2014;64:437–49.

[10] Barbato A, Capone A, Chen L, Martignon F, Paris S. A distributed demand-side management framework for the smart grid. Comput Commun 2015;57:13–24.

[11] Caprino D, Della Vedova ML, Facchinetti T. Peak shaving through real-time scheduling of household appliances. Energy Build 2014;75:133–48.

[12] Vardakas JS, Zorba N, Verikoukis C V. Scheduling policies for two-state smart-home appliances in dynamic electricity pricing environments. Energy 2014;69:455–69.

[13] Iwafune Y, Ikegami T, Fonseca JG da S, Oozeki T, Ogimoto K. Cooperative home energy management using batteries for a photovoltaic system considering the diversity of households. Energy Convers Manag 2015;96:322–9.

[14] Rastegar M, Fotuhi-Firuzabad M, Lehtonen M. Home load management in a residential energy hub. Electr Power Syst Res 2015;119:322–8.

[15] Monacchi A, Zhevzhyk S, Elmenreich W. HEMS: a home energy market simulator. Comput Sci - Res Dev 2014:1–8.

[16] Wang C, Zhou Y, Jiao B, Wang Y, Liu W, Wang D. Robust optimization for load scheduling of a smart home with photovoltaic system. Energy Convers Manag 2015.

[17] Vardakas JS, Zorba N, Verikoukis C V. Performance evaluation of power demand scheduling scenarios in a smart grid environment. Appl Energy 2015;142:164–78.

[18] Mohsenian-Rad AH, Wong VWS, Jatskevich J, Schober R, Leon-Garcia A. Autonomous demand-side management based on game-theoretic energy consumption scheduling for the future smart grid. IEEE Transactions on Smart Grid 2010;1:320–31.

[19] Ozturk Y, Senthilkumar D, Kumar S, Lee G. An Intelligent Home Energy Management System to Improve Demand Response. IEEE Transactions on Smart Grid 2013;4:694–701.

[20] Setlhaolo D, Xia X, Zhang J. Optimal scheduling of household appliances for demand response. Electr Power Syst Res 2014;116:24–8.

[21] Fernandes F, Morais H, Vale Z, Ramos C. Dynamic load management in a smart home to participate in demand response events. Energy Build 2014;82:592–606.

[22] Adika CO, Wang L. Smart charging and appliance scheduling approaches to demand side management. Int J Electr Power Energy Syst 2014;57:232–40.

[23] Soares A, Antunes CH, Oliveira C, Gomes Á. A multi-objective genetic approach to domestic load scheduling

10
in an energy management system. Energy 2013;77:144–52.

[24] Rastegar M, Fotuhi-Firuzabad M. Outage Management in Residential Demand Response Programs. IEEE Transactions on Smart Grid 2015;6:1453–62.

[25] Bera S, Gupta P, Misra S. D2S: Dynamic Demand Scheduling in Smart Grid Using Optimal Portfolio Selection Strategy. IEEE Transactions on Smart Grid 2015;6:1434–42.

[26] Erdinc O, Paterakis NG, Mendes TDP, Bakirtzis AG, Catalao JPS. Smart Household Operation Considering Bi-Directional EV and ESS Utilization by Real-Time Pricing-Based DR. IEEE Transactions on Smart Grid 2015;6:1281–91.

[27] Tang C-J, Dai M-R, Chuang C-C, Chiu Y-S, Lin WS. A load control method for small data centers participating in demand response programs. Futur Gener Comput Syst 2014;32:232–45.

[28] Siano P. Demand response and smart grids—A survey. Renew Sustain Energy Rev 2014;30:461–78.

[29] Shimomura Y, Nemoto Y, Akasaka F, Chiba R, Kimita K. A method for designing customer-oriented demand response aggregation service. CIRP Ann - Manuf Technol 2014;63:413–6.

[30] Horowitz S, Mauch B, Sowell F. Forecasting residential air conditioning loads. Appl Energy 2014;132:47–55.

[31] Xiao Y. Communication and Networking in Smart Grids. CRC Press; 2012.

[32] Kostková K, Omelina Ľ, Kyčina P, Jamrich P. An introduction to load management. Electr Power Syst Res 2013;95:184–91.

[33] Palensky P, Dietrich D. Demand side management: Demand response, intelligent energy systems, and smart loads. IEEE Transactions on Ind. Informatics, 2011;7:381–8.

[34] Li C, Tang S, Cao Y, Xu Y, Li Y, Li J, et al. A new stepwise power tariff model and its application for residential consumers in regulated electricity markets. IEEE Transactions on Power Systems 2013;28:300–8.

[35] Gellings CW, Samotyj M. Smart Grid as advanced technology enabler of demand response. Energy Effic 2013;6:685–94.

[36] Miceli R. Energy Management and Smart Grids. Energies 2013;6:2262–90.

[37] Nikzad M, Mozafari B. Reliability assessment of incentive- and priced-based demand response programs in restructured power systems. Int J Electr Power Energy Syst 2014;56:83–96.

[38] Samadi P, Mohsenian-Rad H, Schober R, Wong VWS. Advanced Demand Side Management for the Future Smart Grid Using Mechanism Design. IEEE Transactions on Smart Grid 2012;3:1170–80.

[39] Ma J, Deng J, Song L, Han Z. Incentive Mechanism for Demand Side Management in Smart Grid Using Auction. IEEE Transactions on Smart Grid 2014;5:1379–88.